\documentclass[11pt]{amsart}

\usepackage{amsmath,amsthm,verbatim}
\usepackage{amsfonts}
\usepackage{amssymb}
\usepackage{amscd}
\usepackage{fancyhdr,pslatex,graphicx,layouts}
\usepackage{hyperref}

\newcommand{\bp}{\begin{problem}}

\newcommand{\ep}{\end{problem}}

\newcommand{\ba}{\begin{answer}}

\newcommand{\ea}{\end{answer}}

\newcommand{\ben}{\renewcommand{\theenumi}{\alph{enumi}}

\renewcommand{\labelenumi}{(\theenumi)}\begin{enumerate}}

\newcommand{\een}{\end{enumerate}}

\newcommand{\Mod}[1]{\ (\mathrm{mod}\ #1)}

\begin{document}

\author[]{Dima Grigoriev}
\address{CNRS, Math\'ematiques, Universit\'e de Lille, 59655, Villeneuve d'Ascq, France}
\email{Dmitry.Grigoryev@univ-lille.fr}

\author[]{Vladimir Shpilrain}
\address{Department of Mathematics, The City  College  of New York, New York,
NY 10031} \email{shpil@groups.sci.ccny.cuny.edu}

\title[RSA and redactable blockchains]
{RSA and redactable blockchains}

\begin{abstract}
A blockchain is redactable if a private key holder (e.g. a central authority) can change any single block without violating integrity of the whole blockchain, but no other party can do that. In this paper, we offer a simple method of constructing redactable blockchains inspired by the ideas underlying the well-known RSA encryption scheme. Notably, our method can be used in conjunction with any reasonable hash function that is used to build a blockchain. Public immutability of a blockchain in our construction is based on the computational hardness of the RSA problem and not on properties of the underlying hash function. Corruption resistance is based on the computational hardness of the discrete logarithm problem.

\end{abstract}

\maketitle


\section{Introduction}

A {\it blockchain} is a distributed database that is used to
maintain a continuously growing list of records, called blocks. Each
block contains a link to the previous (or the next) block. A blockchain is typically
managed by a peer-to-peer network collectively adhering to a
protocol for validating new blocks. By design, blockchains are
inherently resistant to modification of the data. Once recorded, the
data in any given block cannot be altered retroactively without the
alteration of all preceding (or subsequent) blocks and a collusion of the network
majority.

A blockchain can be either public or private. Usually, when people
talk about public blockchains, they mean that  anyone can write
data. A typical example of a public blockchain is {\it bitcoin}. In
contrast, a {\it private} blockchain network is where the
participants are known and trusted: for example, an industry group,
or a military unit, or in fact any private network, big or small. In
particular, what is now called by a popular name {\it The Internet
of things} is a good example of a private network where the
blockchain technology could be very useful. The Internet of things
(IoT) is the inter-networking of physical devices (also referred to
as ``smart devices"), e.g. vehicles, or buildings, or other items
embedded with electronics, software, sensors,  and network
connectivity which enable these objects to collect and exchange
data.  The world of IoT is quickly evolving and growing at an
exponential rate. Experts estimate that the IoT will consist of
about 30 billion objects by 2020.

Concerns have been raised that the Internet of things is being
developed rapidly without appropriate consideration of the profound
security challenges involved. Most of the technical security issues
are similar to those of conventional servers, workstations and
smartphones, but the firewall, security update and anti-malware
systems used for those are generally unsuitable for the much
smaller, less capable, IoT devices. In particular,
computer-controlled devices in vehicles such as brakes, engine,
locks, etc., have been shown to be vulnerable to attackers who have
access to the on-board network. In some cases, vehicle computer
systems are Internet-connected, allowing them to be exploited
remotely.

Blockchain technology would provide at least a partial solution to these security
problems.

\subsection{Immutable and redactable blockchains}
\label{hash_blockchain}

The role of hash functions in a blockchain is similar to that of page numbering in a book. There is, however, an important difference. With books, predictable page numbers make it easy to know the order of the pages. If you ripped out all the pages and shuffled them, it would be easy to put them back into the correct order where the story makes sense. With blockchains, each block references the previous (or the next) block, not by the  block number, but by the block's hash function (the ``fingerprint"), which is smarter than a page number because the fingerprint itself is determined by the contents of the block.

By using a fingerprint instead of a timestamp or a numerical sequence reflecting the block number, one also gets a nice way of validating the data. In any blockchain, you can generate the block fingerprints yourself by using the corresponding hashing algorithm. If the fingerprints are consistent with the data, and the fingerprints join up in a chain, then you can be sure that
the blockchain is internally consistent. There are several ways to securely join blocks in a chain. One of the ways is, informally, as follows. Every block $B_i$ has a prefix, which is the hash (or, more generally, a one-way function) of the fingerprint $H(B_{i-1})$ of the previous block $B_{i-1}$. If anyone wants to meddle with any of the data, they would have to regenerate all the fingerprints from that point forwards and the blockchain will look different.

A blockchain is {\it immutable} if, once data has been written to a blockchain no one, not even a central authority (e.g. a system administrator), can change it. This provides benefits for audit. As a provider of data you can prove that your data has not been altered, and as a recipient of data you can be sure that the data has not been altered. These benefits are useful for databases of financial transactions, for example.

On the other hand, with a private blockchain,  someone with higher privileged access, like a systems administrator, may be able to change the data. So how do we manage the risk of an intruder changing data to his advantage if changing is made easy? The answer to that is provided by {\it  redactable blockchains}; these should involve hash functions with a trapdoor or, more generally, one-way functions with a trapdoor. Trapdoor hash functions are a highly useful cryptographic primitive; in particular, it allows an authorized party to compute a collision with a given hash value, even though the hash function is second pre-image resistant to those who do not know a trapdoor. This property is therefore very useful in application to private blockchains since it makes it possible for an authorized party but not for an intruder to make changes in a blockchain if needed. We give more details in Section \ref{hash_trapdoor}.

The need for a blockchain (even a public one!) to be redactable is well explained in \cite{NYT}: ``That permanence has been vital in building trust in the decentralized currencies, which are used by millions of people. But it could severely limit blockchain’s usefulness in other areas of financial services relied on by billions of people. By clashing with new privacy laws like the ``right to be forgotten" and by making it nearly impossible to resolve human error and mischief efficiently, the blockchain’s immutability could end up being its own worst enemy."

\section{Redactable blockchain structure}
\label{hash_trapdoor}

Recall that a blockchain is {\it immutable} if,  once data has been
written to a blockchain no one, not even a central authority, can
change it. This is achieved by using a hash function $H$ to ``seal"
each  individual block, i.e., each block $B_i$ has a fingerprint
$H(B_i)$, and then connecting blocks in a chain by using another
hash function (or just a one-way function) $G$, as described in our
Section \ref{hash_blockchain}. That way, the blocks $B_i$ become
connected in an immutable blockchain because if somebody tampers
with one of the blocks and changes it, he will have to change all
blocks going forward (or backward), together with their fingerprints, to preserve
consistency of the whole blockchain. This is considered logistically infeasible in
most real-life scenarios.

Originally, blockchains were created to support the bitcoin network,
which is public. Immutability for such a network is crucial. More
recently, as we have pointed out in the Introduction,
with the idea of the Internet of Things gaining momentum, private
networks (small or large) have taken the center stage, and this
creates new challenges. In particular, it is desirable, while
preserving the tampering detection property, to allow  someone with
higher privileged access like a systems administrator or another
authority to be able to change the data or erase (``forget") it \cite{Puddu}. A blockchain that can be
changed like that is called {\it redactable}.

To make a blockchain redactable, {\it trapdoor hash functions} are useful.
Trapdoor hash functions have been considered before (see e.g. \cite{Yang}), but having just any trapdoor hash function is not enough to make a blockchain redactable since the authority who wants
to change a block $B$ usually wants to change it to a {\it particular} block $B'$. A way to make a blockchain redactable was first suggested  in
\cite{Giuseppe}. Recently, \cite{Thyagarajan} claimed the first {\it efficient} redactable public blockchain construction. We also mention {\it chameleon hash functions} \cite{KrawczykRabin} that were recently used \cite{Derler}, \cite{Krenn} in redactable blockchain constructions.

Our approach is focused on private blockchains. It is quite different from \cite{Giuseppe},   \cite{Thyagarajan} and other methods and is  simple and easily implementable. Notably, our method can be used in conjunction with any reasonable hash function that is used to build a blockchain. Public immutability of a blockchain (see Section \ref{immutability}) in our construction is based on the computational hardness of the RSA problem and not on properties of the underlying hash function. Corruption resistance (see Section \ref{Corruption}) is based on the computational hardness of the discrete logarithm problem.

\subsection{A particular blockchain structure we use}\label{general}
There are several possible structures of a redactable blockchain. Our general method should  work with any known structure,
but to make an exposition as clear as possible we choose a very simple structure as follows. Each block $B_i$ will be in 3 parts: a permanent prefix $P_i$, the actual content $C_i$,  and a redactable
suffix $X_i$.  There is also a hash $h_i=H(P_i, C_i)$, where $H$ is a public hash function, and a public one-way function $F$ such that $F(h_i, X_i)=P_{i+1}$. To make such a blockchain redactable, a central authority should have a private key that would allow for replacing $C_i$ with an arbitrary $C'_i$ of his/her choice, so that upon a suitable selection of the new suffix $X'_i$, the equality $F(h'_i, X'_i)=P_{i+1}$ would still hold.

Instead of having the prefix  of $B_{i+1}$ depend on the block $B_i$, one can have the prefix  of $B_{i-1}$ depend on $B_i$, in which case the integrity check will have the form $F(h_i, X_i)=P_{i-1}$. Our method, with minor modification, works with this structure just as well.

We emphasize again that the hash function $H$ and the one-way function $F$ should be public since anyone should be able to create a new block in the chain as well as verify the integrity of the blockchain.

\section{An RSA-based implementation}
\label{RSA}

A particular implementation of the general redactable blockchain structure described above is inspired by  the ideas underlying the well-known RSA encryption scheme \cite{RSA} (see also \cite{Benaloh}, \cite{GM}, \cite{GP} for later developments).

\medskip

\noindent {\bf  Public information:}

 – a large integer $n$, which is a product of two large primes

 – a hash function $H$, e. g. SHA-256. (We emphasize again that our method can be used with any reasonable hash function, so the reader can replace SHA-256 here with his/her favorite hash function.)

\medskip

\noindent {\bf Private information:}

– prime factors of $n = pq$. These $p$ and $q$ should be safe primes
(see e.g. [2]), as in modern implementations of RSA. A safe prime is of the form $2r + 1$, where $r$ is another prime.

\medskip

\noindent {\bf Block structure.} In each block $B_i$, there will be a prefix $P_i$,
the actual content $C_i$ (e. g.  a transaction description), and  
a suffix $X_i$, which is a nonzero  integer modulo $n$.
We also want $X_i$ {\it not} to have order 2, i.e., $X_i^2 \ne 1 \Mod{n}$. Thus, whoever builds a block $B_i$, selects $X_i$ at random on integers between 1  and $n-1$  and then checks if $X_i^2 \ne 1 \Mod{n}$. If $X_i^2 = 1 \Mod{n}$, random selection of $X_i$ is repeated.
Once a proper $X_i$ is selected, a public hash function $H$ (e. g. SHA-256) is applied to concatenation of $P_i$ and $C_i$ to produce $h_i=H(P_i, C_i)$, and  $h_i$ is then converted to an integer $d_i$ modulo $n$.   The prefix $P_{i+1}$ of the next block is then computed as $P_{i+1} = (X_i)^{d_i^2+1} \Mod{n}$.

\medskip

\subsection{Private redactability} Now suppose the central authority, Alice, who is in possession of the private key, wants to change the content of a block $B_i$ from $C_i$ to $C'_i$ but does not want to change any other block. Then Alice computes the hash $h'_i = H(P_i, C'_i)$ and converts it to an integer $d'_i$ modulo $n$. The number  $(d_i'^2+1)$ should be relatively prime to $\phi(n)$, the Euler function of $n$. If it is not, then Alice uses a padding to have $(d_i'^2+1)$ relatively prime to $\phi(n)$. Once it is, Alice finds the inverse $e'_i$ of $(d_i'^2+1)$ modulo $\phi(n)$. Then she computes $X'_i = P^{e'_i}_{i+1} \Mod{n}$. The integrity check now gives: $(X'_i)^{d_i'^2+1} = (P^{e'_i}_{i+1})^{d_i'^2+1} = P_{i+1} \Mod{n}$ because $e'_i (d_i'^2+1) = 1 \Mod{\phi(n)}$ and
$(P_{i+1})^{\phi(n)} = 1 \Mod{n}$.

\subsection{Padding} Note that since $n = pq$, we have $\phi(n) = (p-1)(q-1)$. If $(d_i'^2+1)$ is not relatively prime to $\phi(n)$, this means that either (1) $(d_i'^2+1)$ is even, or (2) $(d_i'^2+1)$ is odd but is divisible by a large prime (recall that $p$ and $q$ are safe primes, i.e., $\frac{p-1}{2}$ and $\frac{q-1}{2}$ are primes). Thus,
our padding is going to be as follows: if $(d_i'^2+1)$ is even, we just add 1 to $d'_i$. If $d_i'^2+1$ is even and  divisible by $\frac{p-1}{2}$ or $\frac{q-1}{2}$ (this is unlikely but possible),
then instead of $d'_i+1$, the padding of  $d'_i$ can be $d'_i-1$. Then $(d'_i-1)^2+1=d_i'^2 -2d_i'+2=(d_i'^2+1)-2(d_i'-1)$. If $2(d_i'-1)$ is divisible by $\frac{p-1}{2}$ or $\frac{q-1}{2}$, then
instead of $d'_i-1$, the padding is going to be $d'_i+3$.  Then $(d'_i+3)^2+1=d_i'^2 +6d_i'+10=(d_i'^2+1)+3(2d_i'+3)$. Both $(d_i'-1)$ and $3(2d_i'+3)$ cannot be divisible by $\frac{p-1}{2}$ or $\frac{q-1}{2}$, but (although very unlikely) one  can be divisible by $\frac{p-1}{2}$ and the other by $\frac{q-1}{2}$. In that case, the padding is going to be $d'_i-3$.


Similarly, if $(d_i'^2+1)$ is odd but is divisible by $\frac{p-1}{2}$ or $\frac{q-1}{2}$ (this is unlikely but possible), the  padding of  $d'_i$ is going to be either $d'_i+2$, or $d'_i-2$, or $d'_i+4$, or $d'_i-4$.

\subsection{Public immutability} \label{immutability} As can be seen from the ``Block structure" paragraph above, a party who would like to change the content of a single block, would have to essentially solve the RSA problem: recover $X$ from $n$,
$X^{d^2+1} \Mod{n}$, and $d$, where $d^2+1$ is relatively prime to $\phi(n)$. This is considered computationally infeasible for an appropriate choice of $n$ and a random $d, ~0 < d < n$.


\subsection{Corruption resistance} \label{Corruption} Another way of unauthorized modification of a block in a blockchain is {\it corruption}, i.e., changing the content of the block to something meaningless. To do that, the intruder can start with a random $X_i$ and then look for a number $d_i$ such that $(X_i)^{d_i^2+1} = P_{i+1} \Mod{n}$, for a given $P_{i+1} (\mod n)$. This mathematical problem that the intruder would have to solve is known as the {\it discrete logarithm problem} and is considered computationally infeasible if $n$ is sufficiently large.
\medskip

\noindent {\bf Two-step attack.} The reason why our blockchain integrity condition is $P_{i+1} = (X_i)^{d_i^2+1}
\Mod{n}$ and not just $P_{i+1} = (X_i)^{d_i} \Mod{n}$ is the following ``two-step" attack. Suppose the integrity condition was $P_{i+1} = (X_i)^{d_i} \Mod{n}$, and suppose the attacker was able to find out that the block $B_{i}$ was changed so that $(X_i)^{d_i} = (X_i')^{d_i'} \Mod{n}$, and that the attacker got a hold of $X_i, X_i', d_i$, and $d_i'$. We will now omit the index $i$ to make the following easier to read.

The attacker can corrupt this block (i.e., change it to something meaningless) as follows. Generically,
$g.c.d.(d, d')=1,$ so we may assume that there are $a, b \in Z_n$ such that $d a + d' b = 1$.
Then $X = ((X')^a X^b)^{d'}$. Now if $X'' = (X')^a X^b$ and $d'' = d'd$, then $(X'')^{d''} = ((X')^a X^b)^{d'd} =
(((X')^a X^b)^{d'})^d = X^d$. Thus, the attacker can find another suffix, $X''$, and $d''$ such that
$(X'')^{d''} = X^d$ and therefore corrupt the block $B_i$.

The reason why this (or similar) attack does not work if the integrity condition is $P_{i+1} = (X_i)^{d_i^2+1} \Mod{n}$ is that
$(X'')^{d''} = X^d$ does not imply $(X'')^{d''^2+1} = X^{d^2+1}$.

\section{Other authenticated data structures}
\label{graphs}

A ``chain", or a path, is the simplest kind of a connected graph. This type is adequate and sufficient for public data structures such as cryptocurrencies, except that in those, occasional ``forks" may exist, in which case the underlying graph is a {\it tree}. Authenticated data structures built on trees were considered before (see e.g. \cite{Merkle}), albeit not in the context of the present paper (i.e., not in terms of redactability).
Here we explain how to make a data structure redactable if the underlying graph has a node of degree greater than 2. The following procedure easily generalizes to an arbitrary underlying graph.

Suppose three blocks $B_1$, $B_2$, and $B_3$ are connected in a chain $B_1 \to B_2 \to B_3$ as usual, but the block $B_2$ is also
connected to another block $B'$, which means that in the underlying graph the node corresponding to the block $B_2$ has degree 3. Suppose now a central authority wants to modify content of the block $B_2$. If she follows our procedure from Section \ref{RSA}, she would have to find a suffix $X'_2$ for the block $B_2$ such that
$(X'_2)^{d_2'^2+1} = P_3$ and at the same time $(X'_2)^{d_2'^2+1} = P'$, where $P'$ is the prefix of the block $B'$. This system of equations will not have a solution if $P_3 \ne P'$. A way around this is introducing an ``intermediate"
block $B_{in}$ between $B_2$ and $B'$. The prefix of $B_{in}$ will be the same as that of $B_3$, i.e.,  equal to $(X'_2)^{d_2'^2+1}$. The content of $B_{in}$ can just indicate that this block is intermediate, i.e., does not have any other function. The suffix $X_{in}$ will be selected following the procedure in Section \ref{RSA}, i.e., so that $X_{in}^{d_{in}^2+1} = P_3$.


\bigskip

\noindent {\it Acknowledgement.} Both authors are grateful to the Hausdorff Research Institute for Mathematics, Bonn for its hospitality during the work on this project.

\bibliographystyle{amsplain}

\end{document}